# Limitations on the use of the Heisenberg picture in quantum optics

J.D. Franson[1] and R.A. Brewster[2]
[1]University of Maryland Baltimore County, Baltimore, MD 21250 USA
[2]University of Maryland College Park, College Park, MD 20740 USA

The Schrodinger and Heisenberg pictures are equivalent formulations of quantum mechanics. Here we use the Schrodinger picture to calculate the decoherence that occurs when an optical Schrodinger cat state is passed through a beam splitter. We find that an exponentially large amount of decoherence can occur even when there is a negligible change in the quadrature operator $\hat{x}(t)$ in the Heisenberg picture. Similar results have been observed in the decoherence produced by an optical amplifier, and we suggest that the usual quadrature operators in the Heisenberg picture do not provide a complete description of the output of optical devices.

The Schrodinger and Heisenberg formulations of quantum mechanics are equivalent[1-5]. The Heisenberg picture is often useful for calculating the time evolution of complicated systems, while the Schrodinger picture sometimes provides the most straightforward way to understand the fundamental properties of a system. In this paper, we use a quasiprobability distribution in phase space to calculate the decoherence that occurs when a Schrodinger cat state passes through a beam splitter. We find that an exponentially large amount of decoherence can occur even when there is a relatively small decrease in the amplitude of the field leaving the beam splitter. In contrast, the usual beam splitter transformation in the Heisenberg picture gives a negligibly small change in the quadrature operator $\hat{x}(t)$ under the same conditions. A similar situation occurs when a cat state is passed through a linear optical amplifier[6]. As a result, we suggest that the usual quadrature operators in the Heisenberg picture do not provide a complete description of the output of optical devices.

Consider a quantum state of light $|\psi_0\rangle$ that is incident on a beam splitter in the input path labelled $A$ in Fig. 1. The other input path labeled $B$ will be assumed to be in the vacuum state containing no photons. The corresponding output modes will be denoted by $A'$ and $B'$. If the reflection coefficient $R$ of the beam splitter is nonzero, then part of the vacuum fluctuations in mode $B$ will be coupled into output mode $A'$. This can be described by a quantum noise operator $\hat{N}$ as discussed in more detail below. In addition, the beam splitter couples part of the amplitude of the input state $|\psi_0\rangle$ into output path $B'$, which leaves some amount of "which-path" information in the environment. We will show that the usual Heisenberg-picture treatment of a beam splitter provides a correct description of the quantum noise $\hat{N}$, but it does not describe the additional decoherence due to the which-path information left in the environment.

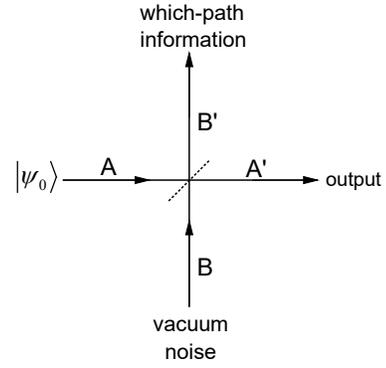

**Figure 1.** A quantum state $|\psi_0\rangle$ is incident on a beam splitter in mode $A$. Vacuum fluctuation noise is coupled from mode $B$ into the output mode $A'$, while which-path information is coupled into the environment in mode $B'$. The Heisenberg operator in Eq. (1) does not include the effects of the which-path information.

We will denote the photon annihilation operator in path $A$ by $\hat{a}$, while the corresponding operator in the other input mode will be denoted by $\hat{b}$. The corresponding operators in the two output modes will be denoted by $\hat{a}'$ and $\hat{b}'$. It can be shown that the input and output quadratures in the Heisenberg picture have the simple property that

$$\hat{x}_A' = T\hat{x}_A - R\hat{\pi}_B = T\hat{x}_A + \hat{N}. \qquad (1)$$

Here $\hat{x}_A = (\hat{a} + \hat{a}^\dagger)/2$ is one of the quadratures of the input electric field in mode $A$, $\hat{x}_A'$ is the corresponding output quadrature, and $\hat{\pi}_B = (\hat{b} - \hat{b}^\dagger)/2i$ is the orthogonal quadrature in input mode $B$. $R$ and $T$ are the reflection and transmission coefficients of the beam splitter, and we have defined the noise operator $\hat{N}$ by $\hat{N} = -R\hat{\pi}_B$. Eq. (1) suggests that the only effect of a beam splitter is to attenuate the amplitude of the field by a factor of $T$ while adding quantum noise $\hat{N}$.

We will focus our attention on the case in which $R \ll 1$. In that limit, $T \rightarrow 1$ while $\hat{N} \rightarrow 0$ and $\hat{x}_A' \rightarrow \hat{x}_A$. In the Heisenberg picture, the output of the beam splitter in Eq. (1) appears to be the same as the input for small values of the reflectivity. Although this may seem intuitively correct, the coherence properties of the output state can be very different from the input, as we will now show.

In order to see this, we will analyze the amount of interference that can occur between the two components of a Schrodinger cat state $|\psi_0\rangle$[7,8] after it has passed through a beam splitter as illustrated in Fig. 2. The cat state of interest is defined by

$$|\psi_0\rangle = c_n \left( |e^{i\phi}\alpha_0\rangle + |e^{-i\phi}\alpha_0\rangle \right), \quad (2)$$

where $c_n$ is a suitable normalizing constant, $\phi$ is a phase shift, and $\alpha_0$ is a complex parameter. A coherent state $|\alpha\rangle$ is defined[9,10] as usual by

$$|\alpha\rangle = e^{-|\alpha|^2/2} \sum_{n=0}^{\infty} \frac{\alpha^n}{\sqrt{n}} |n\rangle, \quad (3)$$

where $|n\rangle$ is a number state of the electromagnetic field containing $n$ photons. The initial cat state corresponds to a superposition of two coherent states with different phases, as illustrated in phase space in Fig. 3, where we have assumed for simplicity that $\alpha_0 = i|\alpha_0|$.

Interference between the two initial components of the cat state can be produced using the interferometer arrangement shown on the right-hand side of Fig. 2[6]. Here a single photon $\gamma$ propagates through an interferometer that contains a Kerr medium K in one of the two paths. Depending on which path the single photon takes, the Kerr medium will apply a phase shift of $\pm\phi$ to the cat state. This will produce an overlap of the two components of the cat state at a phase of $\pi/2$, along with two other non-overlapping probability amplitudes as illustrated in Fig. 3. A homodyne detector is used to measure the phase of the output field, and we only accept (post-select) those events in which the homodyne detector measured a final phase of $\pi/2$. This post-selection process eliminates the contributions from the non-overlapping probability amplitudes, while quantum interference between the two overlapping probability amplitudes will produce a $\cos^2(\theta)$ dependence of the interference pattern. Here $\theta$ is a single-photon phase shift inserted into one of the paths through the interferometer.

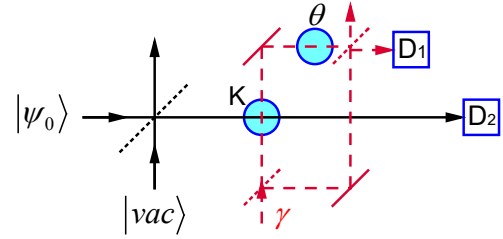

**Figure 2.** Apparatus to measure the amount of quantum interference between the two components of a Schrodinger cat state $|\psi_0\rangle$ after it passes through a beam splitter with the vacuum state $|vac\rangle$ in the other input port. A single photon $\gamma$ passes through an interferometer shown by the red (dashed) lines. A Kerr medium K (along with a constant bias phase shift not shown) will produce a phase shift of $\pm\phi$ depending on the path taken by the photon. $D_1$ is a single-photon detector while $D_2$ is a homodyne detector that measures the phase of the field. The results are post-selected on a single photon detected in $D_1$ with a $90^0$ phase measured in $D_2$. The phase shift of $\pm\phi$ causes two components of the initial cat state to overlap and produce quantum interference that depends on the single-photon phase shift $\theta$, as illustrated in Fig. 3.

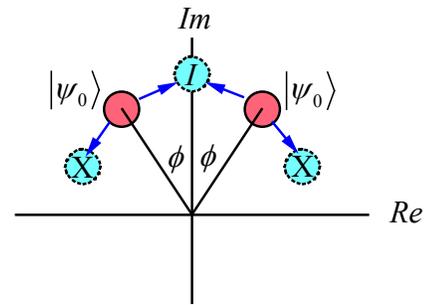

**Figure 3.** Interference of the two components of a cat state produced by the apparatus shown in Fig. 3. The two components of the initial cat state shown in red (solid circles) are displaced by an angle $\phi$ from the imaginary axis in phase space. The Kerr medium in Fig. 2 produces a phase shift of $\pm\phi$ as illustrated by the blue arrows, which displaces the cat state components to the locations indicated by the light blue (dashed) circles. The overlapping components labelled $I$ produce quantum interference, while the non-overlapping components labelled with an $X$ are eliminated by the post-selection process described in Fig. 3.

The visibility of the interference between the two components of the cat state can be analyzed in the

Schrodinger picture using the two-mode Husimi-Kano Q-function[11-13] defined by

$$Q(\alpha',\beta') \equiv \frac{1}{\pi^2}\langle\alpha'|\langle\beta'|\hat{\rho}|\beta'\rangle|\alpha'\rangle. \quad (4)$$

Here $|\alpha'\rangle$ and $|\beta'\rangle$ denote arbitrary coherent states in modes $A'$ and $B'$ of the beam splitter while $\hat{\rho}$ is the density operator for the system. The unitary transformation $\hat{U}$ produced by the beam splitter can be written[14] in the factored form

$$\hat{U} = e^{iR\hat{a}\hat{b}^\dagger/T} T^{\hat{a}^\dagger\hat{a}-\hat{b}^\dagger\hat{b}} e^{iR\hat{a}^\dagger\hat{b}/T}, \quad (5)$$

while the transformation $\hat{V}$ produced by the single-photon interferometer and Kerr cell is given[6] by

$$\hat{V} = \frac{1}{2}\left(e^{i\theta}\hat{\Phi}_+ + \hat{\Phi}_-\right). \quad (6)$$

Here the operators $\hat{\Phi}_\pm = e^{\pm i\phi\hat{n}}$ shift the phase of the field by $\pm\phi$, where $\hat{n}$ is the photon number operator.

The visibility $v$ of the quantum interference pattern can be calculated in the Schrodinger picture using Eqs. (4) through (6), as described in the appendix. The result is that

$$v = e^{-2R^2\sin^2(\phi)|\alpha_0|^2}. \quad (7)$$

It can be seen from Eq. (7) that the visibility will be exponentially small for arbitrarily small values of $R$, provided that the product $R|\alpha_0||\sin(\phi)|$ is much larger than 1. Since $\hat{N} \to 0$ for $R \ll 1$, this shows that the which-path information left in mode $B$ of the beam splitter can seriously degrade the visibility even when the quantum noise $\hat{N}$ is negligible. It is well-known that cat states are very sensitive to photon loss[15, 16], and these results show that the origin of this decoherence is unrelated to the quantum noise operator $\hat{N}$.

The exponential decrease in the visibility in Eq. (7) is inconsistent with the Heisenberg operator in Eq. (1), which seems to imply that the input and output fields should be the same for $R \ll 1$. Although Eq. (1) gives the correct values of the quadrature and its higher moments, it does not describe the entanglement of the field with other modes. In particular, the Heisenberg operator $\hat{x}_A'(t)$ does not describe the loss of coherence in interferometer experiments such as that illustrated in Fig. 2. A similar situation exists for the decoherence of a Schrodinger cat state by an optical amplifier in the Heisenberg picture [6]. These results suggest that the usual quadrature operators in the Heisenberg picture do not provide a complete description of the output of optical devices.

**Acknowledgements**

We would like to acknowledge many valuable discussions with Todd Pittman. This work was supported in part by the National Science Foundation under grant # PHY-1802472.

**Appendix**

Here we outline the calculation of the visibility $v$ in Eq. (7) of the main text in more detail. If we include the second input to the beam splitter, then the initial state $|\psi_0\rangle$ of the system in the Schrodinger picture is

$$|\psi_0\rangle = c_n \left( \left|e^{i\phi}\alpha_0\right\rangle_A + \left|e^{-i\phi}\alpha_0\right\rangle_A \right) \otimes |0\rangle_B, \quad \text{(A1)}$$

where $A$ and $B$ label the two input modes of the beam splitter as in Fig. 2. Here $|0\rangle_B$ denotes the vacuum state in mode B and a coherent state $|\alpha\rangle$ is defined in the text.

Equation (A1) corresponds to a pure state (a Schrodinger cat), whose initial density operator $\hat{\rho}$ can be written in the form[6]

$$\hat{\rho} = \hat{\rho}_{++} + \hat{\rho}_{+-} + \hat{\rho}_{-+} + \hat{\rho}_{--}. \quad \text{(A2)}$$

Here the total density operator $\hat{\rho}$ has been written as the sum of four terms defined by

$$\begin{aligned}
\hat{\rho}_{++} &= c_n^2 \left|e^{i\phi}\alpha_0\right\rangle\left\langle e^{i\phi}\alpha_0\right|_A \otimes |0\rangle\langle 0|_B \\
\hat{\rho}_{+-} &= c_n^2 \left|e^{i\phi}\alpha_0\right\rangle\left\langle e^{-i\phi}\alpha_0\right|_A \otimes |0\rangle\langle 0|_B \\
\hat{\rho}_{-+} &= c_n^2 \left|e^{-i\phi}\alpha_0\right\rangle\left\langle e^{i\phi}\alpha_0\right|_A \otimes |0\rangle\langle 0|_B \\
\hat{\rho}_{--} &= c_n^2 \left|e^{-i\phi}\alpha_0\right\rangle\left\langle e^{-i\phi}\alpha_0\right|_A \otimes |0\rangle\langle 0|_B.
\end{aligned} \quad \text{(A3)}$$

Our goal is to calculate the visibility of the quantum interference in the apparatus shown in Fig. 2 of the main text. This can be done using the two mode Q-function of equation (4). Since the Q-function is a linear function of the density operator, it can be written

$$\begin{aligned}
Q(\alpha', \beta') &= Q_{++}(\alpha', \beta') + Q_{+-}(\alpha', \beta') \\
&+ Q_{-+}(\alpha', \beta') + Q_{--}(\alpha', \beta'),
\end{aligned} \quad \text{(A4)}$$

where $Q_{+-}(\alpha', \beta') = \langle\alpha'|\langle\beta'|\hat{\rho}_{+-}|\beta'\rangle|\alpha'\rangle / \pi^2$, for example.

Eqs. (A2) through (A4) give the initial forms of $\hat{\rho}$ and $Q(\alpha', \beta')$. The final form of $Q(\alpha', \beta')$ can be found by applying the transformation $\hat{V}\hat{U}$ to $\hat{\rho}$. As shown in Fig. 3, we post-select on those situations where the phase shift from the Kerr medium cancels the phase shift in the original cat state component to give a net phase shift of zero. Therefore, we need only keep the relevant term of the operator $\hat{V}$ for each term in the Q-function. For example

$$\begin{aligned}
Q_{+-}(\alpha', \beta') &= \frac{e^{-i\theta}}{\pi^2} \langle\alpha'|_A \langle\beta'|_B \hat{\Phi}_- \hat{U}\hat{\rho}_{+-}\hat{U}^\dagger \hat{\Phi}_+^\dagger |\beta'\rangle_B |\alpha'\rangle_A \\
&= \frac{e^{-i\theta}}{\pi^2} \langle e^{i\phi}\alpha'|_A \langle\beta'|_B \hat{U}\hat{\rho}_{+-}\hat{U}^\dagger |\beta'\rangle_B |e^{-i\phi}\alpha'\rangle_A.
\end{aligned} \quad \text{(A5)}$$

In the last line of Eq. (A5), we have let the operators $\hat{\Phi}_\pm$ (or their adjoints) act on the coherent states to shift their phases. Similar results apply for the other four terms.

We are interested in the interference that results from the apparatus of Fig. 2. The visibility of an interference pattern is defined as

$$v \equiv \frac{P_{max} - P_{min}}{P_{max} + P_{min}}, \quad \text{(A6)}$$

where $P_{max}$ and $P_{\min}$ refer to the maximum and minimum counting rates. It can be shown using Eqs. (A4) through Eq. (A6) that

$$v = \frac{1}{c_n^2} \left| \int Q_{+-}(\alpha', \beta') d^2\alpha' d^2\beta' \right|, \quad \text{(A7)}$$

provided that there was negligible overlap between the two components of the original cat state. Thus, we need only focus on $Q_{+-}(\alpha, \beta)$.

Using Eq. (A3) in Eq. (A5) gives

$$\begin{aligned}
Q_{+-}(\alpha', \beta') &= \frac{c_n^2 e^{-i\theta}}{\pi^2} \left\langle e^{i\phi}\alpha'\right|_A \left\langle\beta'\right|_B \hat{U}|0\rangle_B \left|e^{i\phi}\alpha_0\right\rangle_A \\
&\times \left\langle e^{-i\phi}\alpha_0\right|_A \langle 0|_B \hat{U}^\dagger |\beta'\rangle_B \left|e^{-i\phi}\alpha'\right\rangle_A.
\end{aligned} \quad \text{(A8)}$$

It is convenient [13] to define the variable $f_+$

$$f_+ \equiv \left\langle e^{i\phi}\alpha'\right|_A \langle\beta'|_B \hat{U}|0\rangle_B \left|e^{i\phi}\alpha_0\right\rangle_A, \quad \text{(A9)}$$

with a similar definition of $f_-$. This allows the Q-function of Eq. (A8) to be rewritten as

$$Q_{+-}(\alpha',\beta') = \frac{c_n^2 e^{-i\theta}}{\pi^2} f_+ f_-. \qquad (A10)$$

Inserting Eq. (5) from the main text in Eq. (A9) gives

$$\begin{aligned}
f_+ &= \left\langle e^{i\phi}\alpha' \right|_A \left\langle \beta' \right|_B e^{iR\hat{a}\hat{b}^\dagger/T} T^{\hat{a}^\dagger\hat{a}-\hat{b}^\dagger\hat{b}} e^{iR\hat{a}^\dagger\hat{b}/T} \left|0\right\rangle_B \left|e^{i\phi}\alpha_0\right\rangle_A \\
&= \left\langle e^{i\phi}\alpha' \right|_A \left\langle \beta' \right|_B e^{iR\hat{a}\hat{b}^\dagger/T} T^{\hat{a}^\dagger\hat{a}-\hat{b}^\dagger\hat{b}} \left|0\right\rangle_B \left|e^{i\phi}\alpha_0\right\rangle_A \\
&= e^{-R^2|\alpha_0|^2/2} \left\langle e^{i\phi}\alpha' \right|_A \left\langle \beta' \right|_B e^{iR\hat{a}\hat{b}^\dagger/T} \left|0\right\rangle_B \left|e^{i\phi}T\alpha_0\right\rangle_A \\
&= e^{-R^2|\alpha_0|^2/2} e^{iR\beta'^*\alpha_0 e^{i\phi}} \left\langle e^{i\phi}\alpha' \right|\left.e^{i\phi}T\alpha_0\right\rangle \left\langle \beta'|0\right\rangle.
\end{aligned} \qquad (A11)$$

Here we have made use of the fact that $\hat{b}\left|0\right\rangle_B = 0$ and the transition from the second to third lines can be shown using the number state expansion of a coherent state in Eq. (8).

Using the standard formula for the inner product of two coherent states[11, 12] in Eq. (A11)

$$f_+ = e^{-(|\alpha_0|^2+|\alpha'|^2+|\beta'|^2)/2} e^{T\alpha'^*\alpha_0} e^{iR\beta'^*\alpha_0 e^{i\phi}}, \qquad (A12)$$

with a similar expression for $f_-$. Inserting these values for $f_+$ and $f_-$ into Eq. (A10)

$$\begin{aligned}
Q_{+-}(\alpha',\beta') = &\frac{c_n^2 e^{-i\theta}}{\pi^2} e^{-(|\alpha_0|^2+|\alpha'|^2+|\beta'|^2)} e^{T(\alpha'^*\alpha_0+\alpha_0^*\alpha')} \\
&\times e^{iRe^{i\phi}(\beta'^*\alpha_0-\alpha_0^*\beta')}.
\end{aligned} \qquad (A13)$$

Using Eq. (A13) in Eq. (A7) and performing the integral gives the visibility as

$$v = e^{-2R^2 \sin^2\phi |\alpha_0|^2}, \qquad (A14)$$

which agrees with Eq. (7) of the main text.

We note that there exist simpler methods to arrive at Eq. (A14), but we chose to use the Q-function in order to allow a comparison with previous results for an optical parametric amplifier[13].